\documentclass[letterpaper, 10pt, conference]{IEEEtran}

\IEEEoverridecommandlockouts
\usepackage[letterpaper,left=0.75in,right=0.75in,top=0.75in,bottom=1in]{geometry}

\usepackage{amsmath,amssymb,amsfonts}

\usepackage{amsthm}
\usepackage{booktabs}
\usepackage{tikz}
\usepackage{circuitikz}
\usepackage{xcolor}
\usepackage{url}

\usetikzlibrary{calc,positioning,arrows.meta,fit,backgrounds}

\newtheorem{theorem}{Theorem}
\newtheorem{lemma}{Lemma}
\newtheorem{remark}{Remark}
\newtheorem{problem}{Problem}
\newtheorem{definition}{Definition}

\newcommand{\R}{\mathbb{R}}
\newcommand{\wt}[1]{\tilde{#1}}
\newcommand{\wbar}{\bar{\omega}}

\newcommand{\T}{^{\top}}
\newcommand{\Ibat}{I_{\mathrm{bat}}}
\newcommand{\Lp}{L'}
\newcommand{\Lfp}{L_f'}
\newcommand{\tauref}{\tau_{\mathrm{ref}}}
\newcommand{\ddt}[1]{\tfrac{d #1}{dt}}
\newcommand{\Ecal}{\mathcal{E}}

\begin{document}

\title{\vspace*{0.3in}\LARGE \bf
Passivity Control for Synchronous Machine Systems using Battery, Field Current and Governor Actuation
}

\author{Maryam Khodabakhshloo$^{1}$, Elizabeth L. Ratnam$^{2}$, and Ian R. Petersen$^{1}$
\thanks{$^{1}$M. Khodabakhshloo and I. R. Petersen are with the School of
        Engineering, The Australian National University, Canberra, ACT 2601,
        Australia.}%
\thanks{$^{2}$E. L. Ratnam is with the Department of Electrical and Computer
        Systems Engineering, Monash University, Melbourne, VIC 3800,
        Australia.}%
\thanks{This work was supported by the Australian Research Council under grants DP230102443 and LP210200473.}%
}

\maketitle
\thispagestyle{empty}
\pagestyle{empty}

\begin{abstract}
The passivity of synchronous machines enables power system stability to be established from local conditions. A recent state-feedback design enforces robust passivity using battery actuation at the machine terminal bus. This paper extends that design to a model that includes the field winding and the dynamics of a steam turbine and its governor. The extended model provides two additional actuator signals: the exciter input and the governor input. The nonlinear terms in the model depend on speed and field current deviations, and are represented as bounded uncertainties. Robust positive real conditions are derived in terms of a single algebraic Riccati equation. From its solution, a state-feedback controller is designed in closed form. Numerical results show that, unlike the design with battery actuation alone, the proposed controller leads to passivity at all loading levels when the damping is low.
\end{abstract}

\section{Introduction}

Power systems are moving from a small number of large synchronous generators to many units connected through power electronic converters. This reduces the inertia and damping available in the system, and the remaining synchronous machines can experience larger deviations in speed and current following a disturbance~\cite{Gu2022}. The growing number and variety of units also make it difficult to analyze the stability of the whole network using a single model. Compositional methods address this problem by establishing network stability from the properties of its subsystems~\cite{Arcak2016}. Passivity is a natural property for this purpose, since feedback and parallel interconnections of passive systems are again passive~\cite{Khalil2002}.

The passivity of a synchronous machine, however, is not guaranteed. The port-Hamiltonian stability condition of Caliskan and Tabuada~\cite{Caliskan2014} holds only for small steady-state currents, while our earlier condition~\cite{Khodabakhshloo2025} requires the damping to be large compared with the steady-state current. Using linearized models, Dey and Kulkarni~\cite{Dey2021} show that the real and reactive power controls of a synchronous generator can make it lose passivity below about $10$~Hz. For power systems with Park generator models, Nishino and Ishizaki~\cite{Nishino2025} show that the network subsystem is equilibrium-independent passive only where an energy function is convex. These limitations motivate enforcing passivity by control, using a series compensator~\cite{Caliskan2014}, the field voltage and a series capacitor~\cite{Sun2009}, the shaft torque and the field voltage~\cite{Arghir2018}, or battery based controllers~\cite{Chen2024}. In~\cite{KhodabakhshlooACC}, we designed a battery based state-feedback controller that enforces robust passivity.

The model in~\cite{KhodabakhshlooACC}, however, assumes a constant field current, as in~\cite{Caliskan2014}, and uses the mechanical torque directly as an input, as in~\cite{Arghir2018}. In a real machine, the field current is a state driven by the exciter, and the torque is produced by a turbine whose steam flow is set by the governor valve~\cite{IEEE1973}. Including these dynamics makes the model larger, but it also brings two additional actuators that are already part of the system: the exciter input and the governor input.

In this paper, we use the exciter input and the governor input together with the battery to extend the range of operating conditions over which robust passivity is certified. As in~\cite{KhodabakhshlooACC}, we model the nonlinear terms as bounded uncertainties, which now depend on both the speed and the field current deviations. We then derive robust passivity conditions in the form of an algebraic Riccati equation, and compute the state-feedback gains from its solution. We show that additional actuators can improve the design: a passivity certificate obtained with the battery alone remains valid when the exciter input or the governor input are considered. Numerical results on a benchmark machine show that, when the damping is low, the design with the battery alone fails at most loads, whereas designs with the exciter input or the governor input certify passivity at all loads. Using the exciter input and the governor input together also requires smaller gains than using either of them alone.

The paper is structured as follows. After the preliminaries in Section~\ref{sec:prelim}, Section~\ref{sec:model} presents the model and states the problem, Section~\ref{sec:design} derives the controller, Section~\ref{sec:numerical} gives the numerical results, and Section~\ref{sec:concl} concludes the paper.

\section{Preliminaries}\label{sec:prelim}
\noindent Dissipativity, introduced by Willems~\cite{Willems1972}, describes a system through an energy balance between a \emph{supply rate}, representing the power delivered through an input--output port, and a \emph{storage function}, representing the energy stored in the system. Passivity is a special case of dissipativity in which the supply rate is the product $w\T y$ of the input and output~\cite{Arcak2016}.

\begin{definition}\label{def:passivity}
Consider the system
\begin{equation*}
\ddt{x}=f(x,w),\qquad y=h(x,w),
\end{equation*}
with state $x\in\R^{n}$, input $w\in\R^{m}$, output $y\in\R^{m}$, $f(0,0)=0$ and $h(0,0)=0$. The system is \emph{passive} on a set $\mathcal{X}\subseteq\R^n$ if there exists a continuously differentiable storage function $V:\R^{n}\to\R_{\ge0}$ with $V(0)=0$ such that
\begin{equation}\label{eq:diss-ineq}
\ddt{V}(x)=\frac{\partial V}{\partial x}(x)\,f(x,w)\;\le\; w\T y
\end{equation}
for all $x\in\mathcal{X}$ and all $w\in\R^m$.
\end{definition}

\noindent As long as $x(\tau)\in\mathcal{X}$ for $\tau\in[0,t]$, integrating~\eqref{eq:diss-ineq} gives $V(x(t))-V(x(0))\le\int_0^t w\T y\,d\tau$. So, the increase in stored energy cannot exceed the energy supplied through the port. The storage function only needs to be nonnegative, so it need not depend on every state. In particular, $V(x)=\tfrac12x\T Px$ with $P\succeq0$ is a valid storage function.

\noindent In power systems, the equilibrium is determined by the power flow and is not at the origin. We therefore consider deviations $\wt x=x-x^s$, $\wt w=w-w^s$ and $\wt y=y-y^s$ from a chosen equilibrium $(x^s,w^s,y^s)$. Following the notion of shifted passivity~\cite{vanderSchaft2017}, we apply Definition~\ref{def:passivity} to these deviations, with the storage function measuring the energy relative to the chosen operating point.

\noindent For linear time-invariant systems with a quadratic storage function, the dissipation inequality~\eqref{eq:diss-ineq} reduces to matrix conditions given by the Positive Real Lemma, a special case of the Kalman--Yakubovich--Popov Lemma.  We use this result as the basis for deriving robust passivity conditions for the nonlinear model.

\begin{lemma}[{\cite{AndersonVongpanitlerd1973}}]\label{lem:kyp}
Consider an $m\times m$ transfer matrix $Z(s)=C(sI-A)^{-1}B+D$ with minimal realization $\{A,B,C,D\}$. Then $Z(s)$ is positive real exactly when there exist $P=P\T\succ0$, $L\in\R^{n\times m}$ and $W_0\in\R^{m\times m}$ such that
\begin{equation}\label{eq:kyp}
\begin{aligned}
PA+A\T P&=-LL\T, & PB-C\T&=-LW_0,\\ D+D\T&=W_0\T W_0 .
\end{aligned}
\end{equation}
\end{lemma}

\noindent If~\eqref{eq:kyp} holds with $P=P\T\succeq0$, the system $\ddt{x}=Ax+Bw$, $y=Cx+Dw$ is passive with storage function $V(x)=\tfrac12x\T Px$, even when the realization is not minimal.

\section{System Model and Problem Formulation}\label{sec:model}
\noindent In this section, we derive a model of the synchronous machine system, write its nonlinear terms  as bounded uncertainties, and formulate the robust passivity control problem.

\subsection{Synchronous machine system}
\begin{figure}[t]
\centering
\resizebox{\columnwidth}{!}{%
\begin{circuitikz}[
  every node/.style={font=\large},
  blk/.style={draw,thick,fill=white,minimum height=1.25cm,minimum width=2.1cm,align=center},
  frameM/.style={draw=blue!60!black,dashed,rounded corners,thick,inner sep=0pt},
  frameB/.style={draw=green!45!black,dashed,rounded corners,thick,inner sep=0pt},
  frameG/.style={draw=red!60!black,dashed,rounded corners,thick,inner sep=0pt}]

\coordinate (B) at (5.4,3);   

\draw (0,0) to[sV, l_=SM] (0,3)
      to[R, l=$R_s$] (1.9,3)
      to[L, l=$L$] (3.5,3)
      to[short, i=$i_{abc}$] (4.6,3) -- (B);
\draw (0,0) -- (11.8,0);

\node[draw,circle,thick,minimum size=1.1cm,fill=gray!10] (r) at (-2.3,1.5) {};
\draw[-{Latex[length=2mm]},thick] (r.center) -- ++(40:0.5);
\draw[-{Latex[length=1.6mm]}] ([shift=(70:0.8)]r.center) arc (70:150:0.8);
\draw[densely dotted,thick] (r.east) -- (-0.45,1.5);
\node[left=6pt of r, align=center] {$\theta,\ \omega$\\[1pt]\normalsize $J,\ D_r$};
\draw[-{Latex[length=2.2mm]},thick] (-2.3,4.1) node[above]{$v_f$} -- node[right, align=left]{\normalsize field winding\\[-1pt]\normalsize $R_f,\ L_f,\ i_f$} (r.north);

\fill (B) circle (2.5pt);
\node[above=3pt] at (B) {bus};
\draw (B) -- (11.8,3);
\draw[-{Latex[length=2.2mm]}] (5.9,0.15) -- node[right]{$V_{abc}$} (5.9,2.8);
\draw (7.4,3) to[R, l=$R$] (7.4,0);
\draw (9.1,0) to[american controlled current source, l_=$\Ibat{}_{,abc}$] (9.1,3);
\draw (11.8,0) to[american current source, l_=$I_{abc}$] (11.8,3);
\node[above=3pt] at (11.8,3) {grid};

\node (ref) at (7.0,-2.4) {$\tauref$};
\node[blk] (servo) at (4.3,-2.4) {servomotor\\[2pt]$\dfrac{1}{1+sT_1}$};
\node[blk] (turb)  at (1.0,-2.4) {steam chest\\[2pt]$\dfrac{1}{1+sT_2}$};
\draw[-{Latex[length=2.2mm]},thick] (ref) -- (servo);
\draw[-{Latex[length=2.2mm]},thick] (servo) -- node[above]{$\tau_v$} (turb);
\draw[-{Latex[length=2.2mm]},thick] (turb.west) -- (-2.3,-2.4) -- node[left, pos=0.35]{$T_m$} (r.south);

\begin{scope}[on background layer]
\node[frameM, fit={(-4.2,-0.8) (4.4,5.0)}] (FM) {};
\node[frameB, fit={(4.8,-0.8) (13.4,4.1)}] (FB) {};
\node[frameG, fit={(-1.0,-3.4) (7.7,-1.4)}] (FG) {};
\end{scope}
\node[above=3pt of FM.north, text=blue!60!black] {\textbf{Synchronous machine}};
\node[above=3pt of FB.north, text=green!45!black] {\textbf{Terminal bus}};
\node[below=3pt of FG.south, text=red!60!black] {\textbf{Turbine and governor}};
\end{circuitikz}%
}
\caption{The synchronous machine, with field voltage $v_f$, is connected to a terminal bus with a local resistive load $R$, a storage battery $\Ibat{}_{,abc}$ and the grid, represented by the current $I_{abc}$. The shaft torque $T_m$ is provided by  a non-reheat steam turbine and its governor~\cite{IEEE1973}. The control inputs are $\Ibat{}_{,abc}$, $v_f$ and $\tauref$. Passivity is considered from the grid current $I_{abc}$ to the terminal voltage $V_{abc}$.}
\label{fig:station}
\end{figure}
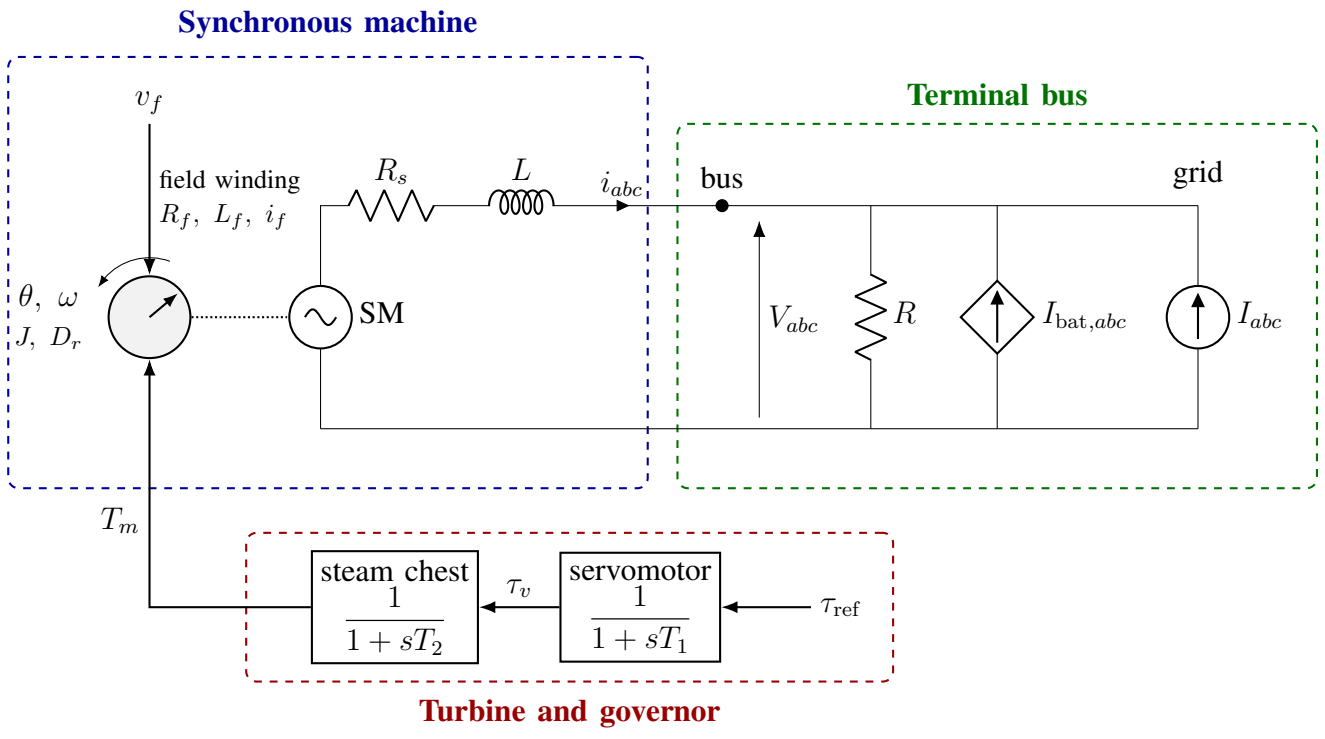

\noindent As shown in Fig.~\ref{fig:station}, a synchronous machine (SM) is connected through the stator and line resistance $R_s$ and inductance $L$ to a bus. The bus is connected to a local resistive load $R$, the grid, represented by a current source $I_{abc}$, and a storage battery, represented by a controllable current source $\Ibat{}_{,abc}$. We assume that the machine has a round rotor with one pole pair, identical, balanced and sinusoidally distributed stator windings, and a single field winding with resistance $R_f>0$ and self-inductance $L_f>0$. Damper windings and magnetic saturation are neglected, and all three-phase quantities are balanced.

\noindent We denote the rotor angle and angular velocity by $\theta$ and $\omega$, the stator currents by $i_{abc}$ and the field current by $i_f$, and define
\begin{equation*}
\mathbf{c}(\theta)=\begin{bmatrix}\cos\theta\\ \cos\bigl(\theta-\tfrac{2\pi}{3}\bigr)\\ \cos\bigl(\theta-\tfrac{4\pi}{3}\bigr)\end{bmatrix}\!,\qquad
\mathbf{s}(\theta)=\begin{bmatrix}\sin\theta\\ \sin\bigl(\theta-\tfrac{2\pi}{3}\bigr)\\ \sin\bigl(\theta-\tfrac{4\pi}{3}\bigr)\end{bmatrix}\!.
\end{equation*}
The stator and field flux linkages are
\begin{equation*}
\lambda_{abc}=L\,i_{abc}+M_f\,\mathbf{c}(\theta)\,i_f,\qquad
\lambda_f=M_f\,\mathbf{c}(\theta)\T i_{abc}+L_f\,i_f ,
\end{equation*}
where $M_f$ is the peak mutual inductance between the stator and field windings and $L$ also includes the line inductance.

\noindent The synchronous machine dynamics follow from Newton's second law, Faraday's law, and Kirchhoff's current law, with the stator currents defined as positive when they leave the machine. Newton's second law for the rotor gives the first two equations in~\eqref{eq:abc}. Faraday's law, applied to the stator and field windings, gives the third and fourth equations. Kirchhoff's current law at the terminal bus gives the last equation:
\begin{equation}\label{eq:abc}
\begin{aligned}
\ddt{\theta} &= \omega,\\
J\ddt{\omega} &= -D_r\,\omega - M_f\,i_f\,\mathbf{s}(\theta)\T i_{abc} + T_m,\\
\ddt{\lambda_{abc}} &= -R_s\,i_{abc} - V_{abc},\\
\ddt{\lambda_f} &= -R_f\,i_f + v_f,\\
V_{abc} &= R\,\bigl(i_{abc}+I_{abc}+\Ibat{}_{,abc}\bigr),
\end{aligned}
\end{equation}
with rotor inertia $J$, damping coefficient $D_r$, shaft torque $T_m$ and field voltage $v_f$. The term $M_f\,i_f\,\mathbf{s}(\theta)\T i_{abc}$ represents the electrical torque, which opposes the shaft torque~\cite{ZhongWeiss2011}.

\noindent The shaft torque is produced by a non-reheat steam turbine. We use a simplified version of the general speed governing and steam turbine models in~\cite[Figs.~4(A) and~7(A)]{IEEE1973}. The model consists of a servomotor lag describing the valve movement, followed by a steam chest lag. Both dynamics are written in torque units:
\begin{equation*}
\begin{aligned}
T_1\ddt{\tau_v} &= -\tau_v + \tauref,\\
T_2\ddt{T_m} &= -T_m + \tau_v,
\end{aligned}
\end{equation*}
where $\tau_v$ is the valve position expressed as a torque and $\tauref$ is the valve command. Dead band and valve rate and position limits are neglected, and the boiler pressure  is assumed  constant.

\subsection{The $dq$ model}
\noindent We apply the power invariant Park transformation $T(\theta)$ with the rotor angle as in reference~\cite{KhodabakhshlooACC}, such that $x_{dq}=T(\theta)x_{abc}$, and define $M:=\sqrt{3/2}\,M_f$. Applying the $dq$ transformation to~\eqref{eq:abc} gives
\begin{equation}\label{eq:dq-raw}
\begin{aligned}
J\ddt{\omega} &= -D_r\,\omega-M\,i_f\,i_q+T_m,\\
\Lp\ddt{i_d} &= -(R_s{+}R)\,i_d-\omega L\,i_q-R\,(I_d+\Ibat{}_{,d})\\
&\quad+\tfrac{M}{L_f}\,(R_f\,i_f-v_f),\\
L\ddt{i_q} &= -(R_s{+}R)\,i_q+\omega L\,i_d+\omega M\,i_f\\
&\quad-R\,(I_q+\Ibat{}_{,q}),\\
\Lfp\ddt{i_f} &= -R_f\,i_f+v_f\\
&\quad+\tfrac{M}{L}\bigl((R_s{+}R)\,i_d+\omega L\,i_q+R\,(I_d+\Ibat{}_{,d})\bigr),\\
V_{dq} &= R\,\bigl(i_{dq}+I_{dq}+\Ibat{}_{,dq}\bigr),
\end{aligned}
\end{equation}
where $\Lp=(LL_f-M^2)/L_f$ and $\Lfp=(LL_f-M^2)/L$. The rotor angle $\theta$ no longer appears in the transformed equations, so its dynamics are omitted.

\noindent As explained in Section~\ref{sec:prelim}, we study passivity for deviations from an equilibrium. Consider an equilibrium~\cite{KhodabakhshlooEquilibria2025} $(\omega^s,i^s_d,i^s_q,i^s_f,\tau^s_v,T^s_m)$ with grid current $I^s_{dq}$, field voltage $v_f^s=R_f i_f^s$, governor input $\tauref^s=T_m^s$ and zero battery current. For each variable $z$, we define its deviation from the equilibrium as $\wt z:=z-z^s$. Subtracting the equilibrium equations from~\eqref{eq:dq-raw} gives the perturbed model. We define the state, grid input, battery input, governor and exciter input, and output as
\begin{equation*}
\begin{aligned}
x&=\begin{bmatrix}\wt\omega\\ \wt i_d\\ \wt i_q\\ \wt i_f\\ \wt\tau_v\\ \wt T_m\end{bmatrix}\!,\;\;
w=\begin{bmatrix}\wt I_d\\ \wt I_q\end{bmatrix}\!,\;\;
u_b=\begin{bmatrix}\wt I_{\mathrm{bat},d}\\ \wt I_{\mathrm{bat},q}\end{bmatrix}\!,\\
u_e&=\begin{bmatrix}\wt\tau_{\mathrm{ref}}\\ \wt v_f\end{bmatrix}\!,\;\;
y=\begin{bmatrix}\wt V_d\\ \wt V_q\end{bmatrix}\!,
\end{aligned}
\end{equation*}
where $w$ is the deviation of the grid current, $u_b$ is the battery input, $u_e$ contains the governor and exciter inputs, and $y$ is the deviation of the terminal voltage. The resulting model is
\begin{equation}\label{eq:pert}
\begin{aligned}
\ddt{x} &= Ax + n(x) + B_w\,w + B_w\,u_b + B_e\,u_e,\\
y &= Cx + D\,w + D\,u_b ,
\end{aligned}
\end{equation}
where the system matrices are given in~\eqref{eq:matrices}, and the nonlinear term is
\begin{equation}\label{eq:nx}
n(x)=\begin{bmatrix}
-\tfrac{M}{J}\,\wt i_f\,\wt i_q\\[2pt]
-\tfrac{L}{\Lp}\,\wt\omega\,\wt i_q\\[2pt]
\wt\omega\,\wt i_d+\tfrac{M}{L}\,\wt\omega\,\wt i_f\\[2pt]
\tfrac{M}{\Lfp}\,\wt\omega\,\wt i_q\\[2pt]
0\\ 0
\end{bmatrix}.
\end{equation}
Since the battery and grid inputs are connected to the same bus, they have the same input matrix $B_w$ and feedthrough matrix $D$. The governor and exciter inputs do not directly affect the terminal voltage.

\begin{figure*}[!t]
\small
\begin{equation}\label{eq:matrices}
\begin{aligned}
A&=\begin{bmatrix}
-\dfrac{D_r}{J} & 0 & -\dfrac{M}{J}i_f^{s} & -\dfrac{M}{J}i_q^{s} & 0 & \dfrac{1}{J}\\[3pt]
-\dfrac{L}{\Lp}i_q^{s} & -\dfrac{R_s{+}R}{\Lp} & -\dfrac{L}{\Lp}\omega^s & \dfrac{MR_f}{L_f\Lp} & 0 & 0\\[3pt]
i_d^{s}+\dfrac{M}{L}i_f^{s} & \omega^s & -\dfrac{R_s{+}R}{L} & \dfrac{M}{L}\omega^s & 0 & 0\\[3pt]
\dfrac{M}{\Lfp}i_q^{s} & \dfrac{M(R_s{+}R)}{L\Lfp} & \dfrac{M}{\Lfp}\omega^s & -\dfrac{R_f}{\Lfp} & 0 & 0\\[3pt]
0 & 0&0&0& -\dfrac{1}{T_1} & 0\\[3pt]
0&0&0&0& \dfrac{1}{T_2} & -\dfrac{1}{T_2}
\end{bmatrix},
\qquad
B_w=\begin{bmatrix}
0&0\\[1pt] -\dfrac{R}{\Lp}&0\\[3pt] 0&-\dfrac{R}{L}\\[3pt]
\dfrac{MR}{L\Lfp}&0\\[3pt] 0&0\\[1pt] 0&0
\end{bmatrix},
\qquad
B_e=\begin{bmatrix}
0&0\\[1pt] 0&-\dfrac{M}{L_f\Lp}\\[3pt] 0&0\\[1pt] 0&\dfrac{1}{\Lfp}\\[3pt] \dfrac{1}{T_1}&0\\[3pt] 0&0
\end{bmatrix},\\[2pt]
C&=\begin{bmatrix}0&R&0&0&0&0\\ 0&0&R&0&0&0\end{bmatrix},
\qquad
D=\begin{bmatrix}R&0\\ 0&R\end{bmatrix},\\[4pt]
B_1&=\begin{bmatrix}
0 & 0 & -\dfrac{M}{J}\\[2pt]
-\dfrac{L}{\Lp} & 0 & 0\\[2pt]
0 & 1 & 0\\[1pt]
\dfrac{M}{\Lfp} & 0 & 0\\[2pt]
0&0&0\\ 0&0&0
\end{bmatrix},
\qquad
C_1=\begin{bmatrix}
0&0&1&0&0&0\\[2pt]
0&1&0&\dfrac{M}{L}&0&0\\[1pt]
0&0&1&0&0&0
\end{bmatrix},
\qquad
\Delta=\begin{bmatrix}\delta_1&0&0\\ 0&\delta_1&0\\ 0&0&\delta_2\end{bmatrix}.
\end{aligned}
\end{equation}
\hrulefill
\vspace*{-2mm}
\end{figure*}

\subsection{Nonlinear terms as a bounded uncertainty}
\noindent The nonlinear term~\eqref{eq:nx} contains the terms $\wt\omega\,\wt i_d$, $\wt\omega\,\wt i_q$, $\wt\omega\,\wt i_f$ and $\wt i_f\,\wt i_q$. In~\cite{KhodabakhshlooACC}, where the field current is  assumed constant, only $\wt\omega\,\wt i_d$ and $\wt\omega\,\wt i_q$ appear. To account for these nonlinearities, we consider $\wt\omega$ and $\wt i_f$ as uncertain parameters with known bounds,
\begin{equation}\label{eq:bounds}
|\wt\omega(t)|\le\wbar,\qquad |\wt i_f(t)|\le \bar{i}_f ,
\end{equation}
where $\wbar$ can be taken as the largest frequency deviation allowed after a credible contingency~\cite{AEMOfreq} and $ \bar{i}_f$ as the admissible variation of the field current. With $\delta_1=\wt\omega$, $\delta_2=\wt i_f$ and the matrices $B_1$, $C_1$ and $\Delta$ as in~\eqref{eq:matrices}, the nonlinear term is written as $n(x)=B_1\Delta C_1x$, and the perturbed model~\eqref{eq:pert} becomes
\begin{equation}\label{eq:lft}
\begin{aligned}
\ddt{x} &= (A+B_1\Delta C_1)\,x + B_w\,w + B_w\,u_b + B_e\,u_e,\\
y &= Cx + D\,w + D\,u_b ,
\end{aligned}
\end{equation}
where $\Delta$ belongs to the set $\boldsymbol{\Delta}$ of matrices of the form in~\eqref{eq:matrices} with $|\delta_1|\le\wbar$ and $|\delta_2|\le\bar{i}_f$, that is, for $x$ in the set $\mathcal{X}=\{x:\ |x_1|\le\wbar,\ |x_4|\le\bar{i}_f\}$.

\subsection{Problem statement}
\noindent Since $\Delta$ is only known to belong to $\boldsymbol{\Delta}$, we adopt the following definition of robust passivity.

\begin{definition}\label{def:robust}
The uncertain system
\begin{equation}\label{eq:gen}
\ddt{x}=(A+B_1\Delta C_1)\,x+B_w\,w,\qquad y=Cx+Dw,
\end{equation}
is \emph{robustly passive} on $\mathcal{X}$ if there exists $P=P\T\succ0$ such that the storage function $V(x)=\tfrac12x\T Px$ satisfies $\ddt{V}\le w\T y$ for all $x\in\mathcal{X}$, $w\in\R^2$ and $\Delta\in\boldsymbol{\Delta}$.
\end{definition}

\begin{problem}[Robust passivity control]\label{prob:design}
Find state-feedback gains $K_b\in\R^{2\times6}$ and $K_e\in\R^{2\times6}$ such that, with $u_b=K_bx$ and $u_e=K_ex$ in~\eqref{eq:lft}, the closed-loop system
\begin{equation}\label{eq:cl}
\begin{aligned}
\ddt{x} &= (A_{\mathrm{cl}}+B_1\Delta C_1)\,x + B_w\,w,\qquad y = C_{\mathrm{cl}}\,x + D\,w,\\
A_{\mathrm{cl}}&=A+B_wK_b+B_eK_e,\qquad C_{\mathrm{cl}}=C+D\,K_b,
\end{aligned}
\end{equation}
is robustly passive on $\mathcal{X}$.
\end{problem}

\section{Robust Passivity Control}\label{sec:design}
\noindent In this section, we first derive conditions for robust passivity of the uncertain system~\eqref{eq:gen}  on $\mathcal{X}$. We bound the uncertainty and use this bound to derive an algebraic Riccati equation. We then apply these conditions to the closed-loop system~\eqref{eq:cl} to design the controller gains for the battery, exciter, and governor inputs.

\subsection{Bounding the uncertainty}
\noindent Along the trajectories of~\eqref{eq:gen}, the derivative of the storage function $V(x)=\tfrac12x\T Px$ includes the uncertain term $\tfrac12x\T(PB_1\Delta C_1+C_1\T\Delta B_1\T P)x$. We introduce one positive design parameter for each independent uncertain parameter to bound this term. For $\epsilon_1,\epsilon_2>0$, we define
\begin{equation*}
\Lambda=\begin{bmatrix}\tfrac{\wbar^2}{\epsilon_1^2}&0&0\\[2pt] 0&\tfrac{\wbar^2}{\epsilon_1^2}&0\\[2pt] 0&0&\tfrac{\bar{i}_f^2}{\epsilon_2^2}\end{bmatrix}\!,\quad
\Ecal=\begin{bmatrix}\epsilon_1^2&0&0\\ 0&\epsilon_1^2&0\\ 0&0&\epsilon_2^2\end{bmatrix}\!.
\end{equation*}

\begin{lemma}\label{lem:bound}
For any $P=P\T$, $\Delta\in\boldsymbol{\Delta}$, $\epsilon_1,\epsilon_2>0$ and $x\in\R^6$,
\begin{multline}\label{eq:bound}
x\T\bigl(PB_1\Delta C_1+C_1\T\Delta B_1\T P\bigr)x\\
\le x\T PB_1\Lambda B_1\T Px + x\T C_1\T\Ecal C_1 x .
\end{multline}
\end{lemma}

\begin{proof}
By completing the square, we have
\[
0\le\bigl\|\Ecal^{1/2}C_1x-\Ecal^{-1/2}\Delta B_1\T Px\bigr\|^2.
\]
Expanding and rearranging gives $x\T\bigl(PB_1\Delta C_1+C_1\T\Delta B_1\T P\bigr)x\le x\T PB_1\Delta\Ecal^{-1}\Delta B_1\T Px+x\T C_1\T\Ecal C_1x$. Since $\Delta\Ecal^{-1}\Delta\preceq\Lambda$ by~\eqref{eq:bounds},  the result follows.
\end{proof}

\subsection{Robust positive real conditions}
\noindent Using Lemma~\ref{lem:bound} together with the Positive Real Lemma, we obtain the following sufficient condition for robust passivity.

\begin{theorem}\label{thm:robust-kyp}
Suppose there exist $\epsilon_1,\epsilon_2>0$, $P=P\T\succ0$, $L\in\R^{6\times2}$ and $W_0\in\R^{2\times2}$ such that
\begin{subequations}\label{eq:rkyp}
\begin{align}
PA+A\T P+PB_1\Lambda B_1\T P+C_1\T\Ecal C_1 &\preceq -LL\T, \label{eq:rk-a}\\
PB_w-C\T &= -LW_0, \label{eq:rk-b}\\
D+D\T &= W_0\T W_0. \label{eq:rk-c}
\end{align}
\end{subequations}
Then~\eqref{eq:gen} is robustly passive on $\mathcal{X}$, with storage function $V(x)=\tfrac12x\T Px$.
\end{theorem}

\begin{proof}
For $x\in\mathcal{X}$, we have $\Delta\in\boldsymbol{\Delta}$. The derivative of $V$ along~\eqref{eq:gen} is
\begin{multline*}
\ddt{V}=\tfrac12x\T\bigl[P(A+B_1\Delta C_1)+(A+B_1\Delta C_1)\T P\bigr]x\\
+x\T PB_w w .
\end{multline*}
Lemma~\ref{lem:bound} and~\eqref{eq:rk-a} give
\begin{multline*}
\tfrac12x\T\bigl[P(A+B_1\Delta C_1)+(A+B_1\Delta C_1)\T P\bigr]x\\
\le\tfrac12x\T\bigl[PA+A\T P+PB_1\Lambda B_1\T P+C_1\T\Ecal C_1\bigr]x\\
\le-\tfrac12x\T LL\T x .
\end{multline*}
From~\eqref{eq:rk-b},
\[
x\T PB_w w=x\T C\T w-x\T LW_0w .
\]
Since $y=Cx+Dw$, \eqref{eq:rk-c} gives
\[
\begin{aligned}
w\T y&=x\T C\T w+\tfrac12w\T(D+D\T)w\\
&=x\T C\T w+\tfrac12w\T W_0\T W_0w .
\end{aligned}
\]
Combining the above relations gives
\begin{multline*}
\ddt{V}-w\T y\le-\tfrac12x\T LL\T x-x\T LW_0w-\tfrac12w\T W_0\T W_0w\\
=-\tfrac12\bigl(L\T x+W_0w\bigr)\T\bigl(L\T x+W_0w\bigr)\le0 ,
\end{multline*}
which is the passivity inequality~\eqref{eq:diss-ineq}.
\end{proof}

\noindent The conditions~\eqref{eq:rkyp} involve three unknown matrices: $P$, $L$ and $W_0$. Since $D+D\T=2R\,I_2$, we can choose $W_0=\sqrt{2R}\,I_2$ and $L=(C\T-PB_w)/\sqrt{2R}$, for which~\eqref{eq:rk-b} and~\eqref{eq:rk-c} hold and $LL\T=\tfrac{1}{2R}(PB_w-C\T)(B_w\T P-C)$. Substituting $LL\T$ into~\eqref{eq:rk-a} and adding a margin $\rho I_6$ gives a condition that depends only on $P$.

\begin{theorem}\label{thm:ol-are}
Suppose that, for given $\epsilon_1,\epsilon_2>0$ and $\rho\ge0$, there exists $P=P\T\succ0$ such that
\begin{equation}\label{eq:ol-are}
P\widetilde A+\widetilde A\T P+P\widetilde QP+\widetilde S=0,
\end{equation}
where
\begin{equation*}
\begin{aligned}
\widetilde A&=A-\tfrac{1}{2R}B_wC,\qquad
\widetilde Q=\tfrac{1}{2R}B_wB_w\T+B_1\Lambda B_1\T,\\
\widetilde S&=\tfrac{1}{2R}C\T C+C_1\T\Ecal C_1+\rho I_6.
\end{aligned}
\end{equation*}
Then~\eqref{eq:gen} is robustly passive on $\mathcal{X}$, with storage function $V(x)=\tfrac12x\T Px$.
\end{theorem}

\noindent For $\rho>0$, \eqref{eq:rk-a} holds with strict inequality. For each fixed pair $(\epsilon_1,\epsilon_2)$, the Riccati equation~\eqref{eq:ol-are} can be solved using the Hamiltonian matrix method~\cite{ZhouDoyleGlover1996}.

\subsection{State-feedback design}
\noindent We now apply Theorem~\ref{thm:ol-are} to the closed-loop system~\eqref{eq:cl}. We first combine the battery, governor, and exciter inputs into a single input vector $u=\begin{bmatrix}u_b\T & u_e\T\end{bmatrix}\T$, where $u_b$ is the battery input and $u_e$ contains the governor and exciter inputs. The corresponding input and feedthrough matrices are
\begin{equation*}
B_u=\begin{bmatrix}B_w&B_e\end{bmatrix},\qquad
D_u=\begin{bmatrix}R&0&0&0\\0&R&0&0\end{bmatrix}.
\end{equation*}
With the state feedback $u=Kx$, where $K=\begin{bmatrix}K_b\T & K_e\T\end{bmatrix}\T$, the closed-loop matrices in~\eqref{eq:cl} are $A_{\mathrm{cl}}=A+B_uK$ and $C_{\mathrm{cl}}=C+D_uK$.
Applying Theorem~\ref{thm:ol-are} to the closed-loop system gives the following Riccati equation:
\begin{multline*}
PA_{\mathrm{cl}}+A_{\mathrm{cl}}\T P+PB_1\Lambda B_1\T P+C_1\T\Ecal C_1+\rho I_6\\
+\tfrac{1}{2R}\bigl(PB_w-C_{\mathrm{cl}}\T\bigr)\bigl(B_w\T P-C_{\mathrm{cl}}\bigr)=0 .
\end{multline*}
Substituting $A_{\mathrm{cl}}$ and $C_{\mathrm{cl}}$ yields
\begin{multline}\label{eq:cl-cond}
PA+A\T P+PB_1\Lambda B_1\T P+C_1\T\Ecal C_1+\rho I_6\\
+\tfrac{1}{2R}\bigl(PB_w-C\T\bigr)\bigl(B_w\T P-C\bigr)\\
+\beta K+K\T\beta\T+K\T NK=0,
\end{multline}
where
\begin{equation*}
\begin{aligned}
\beta&=PB_u-\tfrac{1}{2R}\bigl(PB_w-C\T\bigr)D_u,\\
N&=\tfrac{1}{2R}D_u\T D_u
=\begin{bmatrix}\tfrac R2&0&0&0\\0&\tfrac R2&0&0\\0&0&0&0\\0&0&0&0\end{bmatrix}.
\end{aligned}
\end{equation*}
Since the governor and the exciter inputs have no direct feedthrough, the matrix $N$ is singular. Consequently, $K_e$ enters the last line of~\eqref{eq:cl-cond} only linearly, so this term is unbounded from below and has no minimizer over $K$.
We therefore replace $N$ by the positive definite matrix
\begin{equation*}
\widetilde N=N+E=\begin{bmatrix}\tfrac R2&0&0&0\\0&\tfrac R2&0&0\\0&0&\epsilon_3&0\\0&0&0&\epsilon_4\end{bmatrix}\!,\quad
E=\begin{bmatrix}0&0&0&0\\0&0&0&0\\0&0&\epsilon_3&0\\0&0&0&\epsilon_4\end{bmatrix}\!,
\end{equation*}
where $\epsilon_3>0$ and $\epsilon_4>0$ are design parameters for the governor and the exciter, and choose $K=-\widetilde N^{-1}\beta\T$. Since $K\T NK=K\T\widetilde NK-K\T EK$, this choice gives
\begin{equation*}
\beta K+K\T\beta\T+K\T NK=-\beta\widetilde N^{-1}\beta\T-K\T EK\preceq0 .
\end{equation*}
The resulting design, which solves Problem~\ref{prob:design}, is stated next.

\begin{theorem}\label{thm:design}
Suppose that, for given $\epsilon_1,\epsilon_2,\epsilon_3,\epsilon_4>0$ and $\rho\ge0$, there exists $P=P\T\succ0$ such that
\begin{equation}\label{eq:cl-are}
P\widetilde A_c+\widetilde A_c\T P+P\widetilde Q_cP+\widetilde S_c=0,
\end{equation}
where
\begin{equation*}
\begin{aligned}
\widetilde A_c&=A-\tfrac{1}{R}B_wC,\quad
\widetilde Q_c=B_1\Lambda B_1\T-2B_e\begin{bmatrix}\tfrac{1}{\epsilon_3}&0\\0&\tfrac{1}{\epsilon_4}\end{bmatrix}B_e\T,\\
\widetilde S_c&=C_1\T\Ecal C_1+\rho I_6.
\end{aligned}
\end{equation*}
Then the closed-loop system~\eqref{eq:cl} with the state feedback
\begin{equation}\label{eq:gains}
u_b=-\tfrac1R\bigl(B_w\T P+C\bigr)x,\qquad u_e=-\begin{bmatrix}\tfrac{1}{\epsilon_3}&0\\0&\tfrac{1}{\epsilon_4}\end{bmatrix}B_e\T P\,x,
\end{equation}
is robustly passive on $\mathcal{X}$, with storage function $V(x)=\tfrac12x\T Px$.
\end{theorem}

\begin{proof}
Using the definitions of $B_u$ and $D_u$, we obtain $\beta=\begin{bmatrix}\tfrac12\bigl(PB_w+C\T\bigr)&PB_e\end{bmatrix}$, and since $\widetilde N$ is diagonal, the gain $K=-\widetilde N^{-1}\beta\T$ gives~\eqref{eq:gains}. With this $K$,
\begin{multline*}
\beta K+K\T\beta\T+K\T NK=-\beta\widetilde N^{-1}\beta\T-K\T EK\\
=-\tfrac{1}{2R}\bigl(PB_w+C\T\bigr)\bigl(B_w\T P+C\bigr)\\
-2PB_e\begin{bmatrix}\tfrac{1}{\epsilon_3}&0\\0&\tfrac{1}{\epsilon_4}\end{bmatrix}B_e\T P .
\end{multline*}
By substituting this into~\eqref{eq:cl-cond} and using
\begin{multline*}
\bigl(PB_w-C\T\bigr)\bigl(B_w\T P-C\bigr)-\bigl(PB_w+C\T\bigr)\bigl(B_w\T P+C\bigr)\\
=-2\bigl(PB_wC+C\T B_w\T P\bigr),
\end{multline*}
we obtain~\eqref{eq:cl-are}. Hence $P$ satisfies the Riccati equation of Theorem~\ref{thm:ol-are} for the closed-loop realization, and Theorem~\ref{thm:ol-are} gives the result.
\end{proof}

\begin{remark}\label{rem:monotone}
Theorems~\ref{thm:ol-are} and~\ref{thm:design} remain valid if the equalities~\eqref{eq:ol-are} and~\eqref{eq:cl-are} are replaced by the inequalities $\preceq0$. Since the last term of $\widetilde Q_c$ is negative semidefinite, adding the valve or the exciter, or decreasing $\epsilon_3$ or $\epsilon_4$, can only enlarge the set of pairs $(\epsilon_1,\epsilon_2)$ for which this inequality has a solution $P\succ0$, at the cost of larger governor and exciter gains.
\end{remark}

\section{Numerical Study}\label{sec:numerical}
\subsection{Machine data and operating points}
\noindent We consider the $160$~MVA, $15$~kV, $60$~Hz generator from~\cite[Example~4.1]{Anderson2002}, including the resistance and self-inductance of the field winding, and use the network connection described in~\cite{Barabanov2017}. The model of the turbine and governor has a servomotor time constant of $T_1=0.1$~s and a steam chest time constant of $T_2=0.3$~s. Both values lie within the typical ranges for a non-reheat unit reported in Tables~I and~II of~\cite{IEEE1973}. The machine, network, and design parameters are listed in Table~\ref{tab:params}.

\begin{table}[!t]
\centering
\caption{Machine, network and design parameters.}
\label{tab:params}
\setlength{\tabcolsep}{4pt}
\begin{tabular}{lll}
\toprule
\textbf{Symbol} & \textbf{Description} & \textbf{Value}\\
\midrule
$\omega^s$ & synchronous speed & $377$ rad/s\\
$R_s$, $L$ & stator and line & $29.7$ m$\Omega$, $7.833$ mH\\
$M$ & $\sqrt{3/2}\,M_f$ & $109.01$ mH\\
$R_f$, $L_f$ & field winding & $0.371\ \Omega$, $2.189$ H\\
$\Lp$, $\Lfp$ & see~\eqref{eq:dq-raw} & $2.404$ mH, $0.672$ H\\
$J$ & inertia & $5336$ kg$\cdot$m$^2$\\
$D_r$ & nominal damping & $2252$ N$\cdot$m$\cdot$s/rad\\
$T_1$, $T_2$ & servomotor, steam chest & $0.1$ s, $0.3$ s\\
$\wbar$, $\bar{i}_f$ & bounds~\eqref{eq:bounds} & $3.14$ rad/s, $0.1\,i_f^s$\\
$\rho$ & margin & $10^{-6}$\\
\bottomrule
\end{tabular}
\end{table}

\noindent We study four operating points at the rated terminal voltage, $\|V^s_{dq}\|=15$~kV, and a lagging power factor of $0.85$. The stator current ranges from $15\%$ to $100\%$ of its rated value of $10\,667$~A. At each operating point, the local resistive load $R$ absorbs $10\%$ of the generator's active power. At steady state, $v_f^s=R_fi_f^s$. The steady-state equations obtained from~\eqref{eq:dq-raw} are
\begin{equation*}
\begin{aligned}
V_d^s&=-R_s\,i_d^s-\omega^sL\,i_q^s,\\
V_q^s&=-R_s\,i_q^s+\omega^sL\,i_d^s+\omega^sM\,i_f^s.
\end{aligned}
\end{equation*}
These equations determine the field current $i_f^s$ and the load angle $\delta$. The resulting operating points are given in Table~\ref{tab:op}.

\noindent For each set of actuators, we fix $\epsilon_3$ and $\epsilon_4$ and solve~\eqref{eq:cl-are} over a logarithmic $26\times31$ grid of $(\epsilon_1,\epsilon_2)$. We call a pair feasible if~\eqref{eq:cl-are} admits a solution $P\succ0$. Among the feasible pairs, we select the one that gives the smallest norm of the state-feedback gain. We use bisection to determine the largest uncertainty bound $\wbar_{\max}$ for which a feasible pair exists.
\begin{table}[!t]
\centering
\caption{Operating points at rated voltage and power factor $0.85$.}
\label{tab:op}
\setlength{\tabcolsep}{3.5pt}
\begin{tabular}{lrrrrrr}
\toprule
Load & $i_d^s$ (A) & $i_q^s$ (A) & $i_f^s$ (A) & $\delta$ & $R$ ($\Omega$) & $P$ (MW)\\
\midrule
$15\%$  & $-1124$  & $1139$ & $437.5$  & $12.8^\circ$ & $110.3$ & $20.4$\\
$30\%$  & $-2572$  & $1904$ & $525.3$  & $21.7^\circ$ & $55.1$  & $40.8$\\
$50\%$  & $-4682$  & $2555$ & $655.6$  & $29.6^\circ$ & $33.1$  & $68.0$\\
$100\%$ & $-10\,124$ & $3358$ & $1010.1$ & $39.9^\circ$ & $16.5$ & $136.0$\\
\bottomrule
\end{tabular}
\end{table}

\subsection{Battery feedback at nominal damping}
\noindent We first consider the battery as the only actuator, that is, $u_e\equiv0$, with the nominal damping $D_r$. For each loading level of Table~\ref{tab:op}, Table~\ref{tab:battery} reports the selected pair $(\epsilon_1,\epsilon_2)$ and the slowest closed-loop pole. A feasible design is obtained  for all four operating points, so the proposed condition certifies passivity with the battery as the only actuator. At rated load, the design remains feasible up to $\wbar_{\max}=8.05$~rad/s, that is, for frequency deviations of $\pm1.28$~Hz.

\begin{table}[!t]
\centering
\caption{Design with the battery alone, nominal damping.}
\label{tab:battery}
\setlength{\tabcolsep}{4pt}
\begin{tabular}{lcc}
\toprule
Load & $(\epsilon_1,\epsilon_2)$ & slowest pole\\
\midrule
$15\%$  & $(0.0251,\,0.0010)$ & $-0.323$\\
$30\%$  & $(0.0251,\,0.0010)$ & $-0.283$\\
$50\%$  & $(0.0158,\,0.0251)$ & $-0.248$\\
$100\%$ & $(0.0158,\,0.0251)$ & $-0.181$\\
\bottomrule
\end{tabular}
\end{table}

\subsection{Exciter and valve feedback at low damping}
\noindent Lower damping is expected in power systems with a high penetration of units connected through power electronic converters~\cite{Gu2022}. We examine how the additional actuators affect passivity certification at lower damping levels. We consider four reduced damping values, $D_r/2$, $D_r/4$, $D_r/8$ and $D_r/16$, and compare four sets of actuators: the battery alone, the battery with the governor, the battery with the exciter, and the battery with both. In each case, we remove the unused inputs from $u_e$ and the corresponding columns of $B_e$ and rows and columns of $E$. We use $\epsilon_3=10^{-8}$ for the governor and $\epsilon_4=1$ for the exciter to account for the different units of their inputs. In particular, the governor input is a torque measured in N$\cdot$m.

\noindent The design with the battery alone becomes less effective as the damping decreases. At $D_r/2$, passivity is still certified at every loading level. At $D_r/4$, no feasible design is found at rated load. At $D_r/8$, passivity is certified only at $15\%$ load. At $D_r/16$, no feasible design is found at any loading level. In contrast, adding either the exciter or the governor allows passivity to be certified at every loading level, even at $D_r/16$. This is consistent with Remark~\ref{rem:monotone} and shows the benefit of the additional actuators when the machine damping is low.

\noindent Since the inputs have different units, we compare the exciter and the valve gains in per unit on the machine base. At rated load with $D_r/16$, the norm of the exciter gain is $1.43$~pu when the exciter is used without the valve, and the norm of the valve gain is $22.4$~pu when the valve is used without the exciter. The smaller exciter gain reflects its direct action on the field current, which is coupled to the $d$ axis current through $M$. When both actuators are used, both gain norms decrease: the norm of the exciter gain decreases to $1.07$~pu and that of the governor gain to $9.2$~pu. Thus, compared with the design with the battery alone in~\cite{KhodabakhshlooACC}, adding the exciter or the governor extends passivity certification to operating points where the battery alone is insufficient. Using both actuators also reduces the required gain of each.

\section{Conclusion}\label{sec:concl}
\noindent In this paper, we extended the robust passivity control approach of~\cite{KhodabakhshlooACC}, which used only a battery actuator, to a synchronous machine system with the field winding and the dynamics of a steam turbine and its governor. In this model, the nonlinear terms depend on two variables, the speed and the field current deviations, and we modeled both of them as bounded uncertainties. We then designed a state-feedback controller for the battery, the exciter and the governor inputs, with gains computed from an algebraic Riccati equation. We showed that the exciter and the governor can improve the design obtained with the battery alone. A numerical study with low damping illustrated this result: the design with the battery input alone failed at most loads, while the design with the exciter input or the governor input certified passivity at all loads.

\noindent In future work, we will study the stability of multimachine networks in which each synchronous machine uses the proposed controller, and then add units connected through power electronic converters to these networks.


\end{document}